\documentclass[12pt]{article}

\usepackage{bbm}
\usepackage{amssymb,amsmath,amsthm}
\usepackage{textcomp}

\usepackage[dvips,xdvi]{graphics}
\usepackage[dvips,xdvi]{graphicx}

\newcommand{\bone}{\mathbbm{1}}

\newcommand{\lang}{\langle\langle}
\newcommand{\rang}{\rangle\rangle}

\newcommand{\be}{\begin{equation}}
\newcommand{\ee}{\end{equation}}
\newcommand{\ba}{\begin{eqnarray}}
\newcommand{\ea}{\end{eqnarray}}

\newcommand{\s}{\hspace{0.5mm}}

\newcommand{\bmat}{\begin{pmatrix}}
\newcommand{\emat}{\end{pmatrix}}
\newcommand{\zed}{\mathbbm{Z}}

\allowdisplaybreaks

\swapnumbers 
\theoremstyle{definition} 
\newtheorem{define}{Definition}[section]

\theoremstyle{definition} 
\newtheorem{theorem}[define]{Theorem}

\theoremstyle{definition}

\theoremstyle{definition} 
\newtheorem{prop}[define]{Proposition}

\begin{document}

\title{
\normalsize \hfill UWThPh-2011-3 \\[12mm]
\LARGE Comments on the classification\\ of the finite subgroups of SU(3)
}

\author{
Patrick Otto Ludl\thanks{E-mail: patrick.ludl@univie.ac.at}
\\*[3mm]
\small University of Vienna, Faculty of Physics \\
\small Boltzmanngasse 5, A--1090 Vienna, Austria
}

\date{}

\maketitle

\begin{abstract}
Many finite subgroups of SU(3) are commonly used in particle physics. The classification of the finite subgroups of SU(3) began with the work of H.F.~Blichfeldt at the beginning of the 20\textsuperscript{th} century. In Blichfeldt's work the two series (C) and (D) of finite subgroups of SU(3) are defined. While the group series $\Delta(3n^2)$ and $\Delta(6n^2)$ (which are subseries of (C) and (D), respectively) have been intensively studied, there is not much knowledge about the group series (C) and (D).
In this work we will show that (C) and (D) have the structures $\mathrm{(C)}\cong(\zed_m\times \zed_{m'})\rtimes\zed_3$ and $\mathrm{(D)}\cong(\zed_{n}\times\zed_{n'})\rtimes S_3$, respectively. Furthermore we will show that, while the (C)-groups can be interpreted as irreducible representations of $\Delta(3n^2)$, the (D)-groups can in general not be interpreted as irreducible representations of $\Delta(6n^2)$.
\end{abstract}

\section{Introduction}\label{introduction}
Today finite groups are widely used in physics, and particle physics offers a wide range of applications for the theory of finite groups in particular. Especially the finite subgroups of SU(3) have been intensively studied in the past, and their investigation and application in various fields of particle physics continues unabated.

The systematic analysis of finite subgroups of SU(3) for application in particle physics started with the work of Fairbairn, Fulton and Klink~\cite{finitesubgroups-su3-fairbairn} in 1964. Since then many contributions to a systematic analysis of these groups for application in particle physics have been published. The finite subgroups of SU(3) have been used for model building in hadron physics~\cite{finitesubgroups-su3-fairbairn}, as well as a computational tool in lattice QCD (see e.g.~\cite{lattice}). Today's most important application is flavour physics: On the one hand there is an enormous amount of models using finite subgroups of SU(3) (see e.g.~\cite{DTPOL,discretesymmreviews} and references therein) trying to solve the fermion mass and mixing problems in the lepton sector as well as the quark sector. On the other hand finite subgroups of SU(3) have also been used in the context of minimal flavour violation (see e.g.~\cite{Zwicky}).

The well-known group series $T_n,\s \Delta(3n^2)$ and $\Delta(6n^2)$ are sub-series of the series (C) and (D)~\cite{miller} of finite subgroups of SU(3). However, (C) and (D) also contain other groups, which have been paid much less attention. The aim of this work is to study the structure of the finite subgroups of SU(3) of type (C) and (D) and their relation to other finite subgroups of SU(3).

\subsection*{The classification of the finite subgroups of SU(3)}
Here we want to list the main efforts that have been put in the classification of the finite subgroups of SU(3) in chronological order.
\begin{itemize}
	\item In 1916 G.A.~Miller, H.F.~Blichfeldt and L.E.~Dickson published their book \textit{Theory and Applications of Finite Groups}~\cite{miller}. In the part written by H.F.~Blichfeldt the finite subgroups of SU(3) are classified in terms of their generators. The series $\Delta(3n^2)$ and $\Delta(6n^2)$ are not explicitly defined but are contained in the series (C) and (D).
	\item In 1964 the article \textit{Finite and Disconnected Subgroups of SU(3) and their Application to the Elementary-Particle Spectrum} by W.M.~Fairbairn, T.~Fulton and W.H.~Klink was published~\cite{finitesubgroups-su3-fairbairn}. It was the first paper which faced the task of analyzing a large set of finite subgroups of SU(3) for their use as symmetries in particle physics (hadron physics in this special case). In~\cite{finitesubgroups-su3-fairbairn} the group series $\Delta(3n^2)$ and $\Delta(6n^2)$ are already included.
	\item In their articles \textit{Representations and Clebsch-Gordan coefficients of $Z$-metacyclic groups}~\cite{BLW1} and \textit{Finite subgroups of SU(3)}~\cite{BLW2} (both published in 1981) A.~Bovier, M.~L\"uling and D.~Wyler defined and analyzed the SU(3)-subgroups $T_n$. Especially they constructed all irreducible representations and calculated all Clebsch-Gordan coefficients for these groups. Furthermore Bovier et al. investigated the group series $\Delta(3n^2)$ and $\Delta(6n^2)$ in detail giving not only the irreducible representations but also the Clebsch-Gordan coefficients for both series.
	\item Two years later, in their paper \textit{Some comments on finite subgroups of SU(3)}~\cite{Fairbairn2}, W.M.~Fairbairn and T.~Fulton proved that some groups of the type $T_n$ given by Bovier et al. in~\cite{BLW2} are not subgroups of SU(3). 
	\item In 2007 C.~Luhn, S.~Nasri and P.~Ramond published their work \textit{The Flavor Group $\Delta(3n^{2})$}~\cite{D3n^2}, giving all conjugacy classes, irreducible representations, character tables and Clebsch-Gordan coefficients of $\Delta(3n^2)$.
	\item In 2008 J.A.~Escobar and C.~Luhn published their analysis \textit{The Flavor Group $\Delta(6n^2)$}~\cite{D6n^2}, giving all conjugacy classes, irreducible representations, character tables and Clebsch-Gordan coefficients of $\Delta(6n^2)$.
	\item In 2009 the work \textit{Systematic analysis of finite family symmetry groups and their application to the lepton sector}~\cite{DTPOL} was published. It contains an analysis and summary of all finite subgroups of SU(3). With the help of~\cite{D3n^2} it could be shown that all SU(3)-subgroups of type (C) can be interpreted as three-dimensional irreducible representations of $\Delta(3n^2)$. The generators of the group series (D) were determined explicitly.\\
In the same year R.~Zwicky and T.~Fischbacher showed that every (D)-group is a subgroup of $\Delta(6n^2)$ for a suitable $n$ in their article \textit{On discrete Minimal Flavour Violation}~\cite{Zwicky}.
	\item In the work \textit{On the finite subgroups of U(3) of order smaller than 512}~\cite{U3-512} (published in 2010) all finite subgroups of U(3) of order smaller than 512 which possess a faithful three-dimensional irreducible representation are listed.\footnote{The list in~\cite{U3-512} contains only those groups which are not isomorphic to a group of the form $H\times\zed_n$ with $n>1$.} Among these groups there is no SU(3)-subgroup which does not fit into the classification scheme of Blichfeldt~\cite{miller}.\\
In their article \textit{Tribimaximal Mixing From Small Groups}~\cite{parattu-wingerter} K.M.~Parattu and A.~Wingerter began to analyze also those finite subgroups of U(3) which possess a faithful three-dimensional \textit{reducible} representation but which do not possess any faithful irreducible representation. Their analysis of all groups up to order 100 shows no finite subgroups of SU(3) which do not fit into Blichfeldt's classification scheme~\cite{miller}.
\end{itemize}
Table~\ref{types-SU3-subgroups} shows the different types of finite subgroups of SU(3), as they are classified by now. Especially it contains all finite subgroups of SO(3), see~\cite{ReviewGrimusLudl,Hamermesh}, as follows:
\begin{itemize}
 \item The \textit{uniaxial groups} (groups of rotations about one axis) are cyclic and thus isomorphic to $\zed_n=A(n,1)$.
 \item The \textit{dihedral groups} possess faithful two-dimensional representations and are thus contained in $B$.
 \item The \textit{tetrahedral group} is isomorphic to $A_4\cong\Delta(12)$.
 \item The \textit{octahedral group} is isomorphic to $S_4\cong\Delta(24)$.
 \item The \textit{icosahedral group} is isomorphic to $A_5\cong\Sigma(60)$. 
\end{itemize}
Only two of the groups presented in table~\ref{types-SU3-subgroups}
are not contained in the list given by Blichfeldt in \cite{miller}, namely the direct products
$\Sigma(60)\times\zed_3$ and $\Sigma(168)\times\zed_3$.

\begin{table}[t]
\begin{center}
\begin{tabular}{|lll|}
\hline
Group & Order & References \\
\hline
$A(m,n)\cong \zed_{m}\times\zed_{n}$ (Abelian groups), $n$ divides $m$ & $mn$ & \\
$B$ (finite subgroups of U(2)) & no general formula & \cite{miller,ReviewGrimusLudl}\\
$C(n,a,b)$ & no general formula & \cite{DTPOL,miller} \\ 
$D(n,a,b;d,r,s)$ & no general formula & \cite{DTPOL,Zwicky,miller}\\
$\Delta(3n^2)\cong C(n,0,1),\enspace n\ge 2$ & $3n^2$ & \cite{finitesubgroups-su3-fairbairn,DTPOL,BLW1,D3n^2}\\
$\Delta(6n^2)\cong D(n,0,1;2,1,1),\enspace n\ge 2$ & $6n^2$ & \cite{finitesubgroups-su3-fairbairn,DTPOL,BLW1,D6n^2}\\
$T_n\cong C(n,1,a),\enspace (1+a+a^2)\s \mathrm{mod}\s n=0$, or & $3n$ & \cite{DTPOL,BLW1,BLW2,Fairbairn2}\\
$T_n\cong C(3p,1,a),\enspace (1+a+a^2)\s \mathrm{mod}\s 3p=0$; $n=3p$ & $3n=9p$ & \cite{DTPOL,BLW1,BLW2,Fairbairn2}\\
$\Sigma(60)\cong A_5$ & 60 & \cite{finitesubgroups-su3-fairbairn,DTPOL,miller,SimpleFinite,A5}\\
$\Sigma(60)\times\zed_3$ & 180 & \\
$\Sigma(168)\cong PSL(2,7)$ & 168 & \cite{finitesubgroups-su3-fairbairn,DTPOL,miller,SimpleFinite}\\
$\Sigma(168)\times\zed_3$ & 504 & \\
$\Sigma(36\times 3)$ & 108 & \cite{finitesubgroups-su3-fairbairn,DTPOL,miller,Bovier-Wyler,Principal_series}\\
$\Sigma(72\times 3)$ & 216 & \cite{finitesubgroups-su3-fairbairn,DTPOL,miller,Bovier-Wyler,Principal_series}\\
$\Sigma(216\times 3)$ & 648 & \cite{finitesubgroups-su3-fairbairn,DTPOL,miller,Bovier-Wyler,Principal_series}\\
$\Sigma(360\times 3)$ & 1080 & \cite{finitesubgroups-su3-fairbairn,DTPOL,miller}\\
\hline
\end{tabular}
\caption{Types of finite subgroups of SU(3)~\cite{finitesubgroups-su3-fairbairn,miller,BLW2,Fairbairn2}. The allowed values for $n$ and $p$ in $T_n$ are products of powers of primes of the form $3k+1$, $k\in\mathbbm{N}$~\cite{BLW2,Fairbairn2}.}
\label{types-SU3-subgroups}
\end{center}
\end{table}
\hspace{0mm}\\

\section{Abelian subgroups of SU(3)}
In the following sections we will frequently deal with Abelian subgroups of SU(3). The remarkably simple theorem~\ref{AbelianSU3subgroups} provides us with all necessary information we will need in our later analysis.

\begin{theorem}\label{AbelianSU3subgroups}
Every finite Abelian subgroup $\mathcal{A}$ of SU(3) is isomorphic to $\zed_m\times\zed_n$, where 
	\be
	m=\max_{a\in \mathcal{A}}\mathrm{ord}(a)
	\ee
and $n$ is a divisor of $m$. The proof of this theorem can be found in appendix~\ref{proofappendix}. 
\end{theorem}

\section{On the SU(3)-subgroups of type (C)}\label{Cgroups}
In this section we will investigate the structure of the SU(3)-subgroups of type (C). Knowing the structure of (C) we can easily show that there exist SU(3)-subgroups of type (C) which neither belong to the series $\Delta(3n^2)$, nor to the groups of type~\cite{BLW1,BLW2,Fairbairn2}
	\[
	T_n=\zed_n\rtimes\zed_3.
	\]
In the following the symbol $\lang ...\rang$ means ``generated by''. The group series (C) is generated by the matrices
	\be
	E:=\begin{pmatrix}
	0 & 1 & 0\\
	0 & 0 & 1\\
	1 & 0 & 0\\	
	\end{pmatrix}
	\quad
	\mbox{and}
	\quad
	F(n,a,b):=\begin{pmatrix}
	\eta^a & 0 & 0\\
	0 & \eta^b & 0\\
	0 & 0 & \eta^{-a-b}\\	
	\end{pmatrix},
	\ee
with $\eta=\exp(2\pi i/n)$.
	\be\label{Cdefinition}
	C(n,a,b):=\lang E,\enspace F(n,a,b)\rang,\quad n\in\mathbbm{N}\backslash \{0\},\enspace a,b\in\{0,...,n-1\}.
	\ee
Since the irreducible three-dimensional representations of $\Delta(3n^2)$ are~\cite{D3n^2}
	\be
	\textbf{\underline{3}}_{(a,b)}:\enspace G_1\mapsto E,\enspace G_2\mapsto F(n,b,a),
	\ee
where $G_1$ and $G_2$ denote the generators of $\Delta(3n^2)$, we find
	\be\label{CDelta3n^2}
	C(n,a,b)\cong\textbf{\underline{3}}_{(b,a)}(\Delta(3n^2)).
	\ee
There is no general formula for the order of $C(n,a,b)$, but we can give a prescription for the calculation of the order of $C(n,a,b)$ for given $n,a,b$.

Let us first think about the structure of $C(n,a,b)$. Defining
	\be
	X:=F(n,a,b),\quad Y:=F(n,b,-a-b),
	\ee
we find the commutation relations
	\be
	XE=EX^{-1}Y^{-1},\quad YE=EX.
	\ee
Therefore
	\begin{itemize}
	\item The subgroup $\lang X,\s Y \rang$ of all diagonal matrices is a normal subgroup of $C(n,a,b)$.
	\item Every element of $C(n,a,b)$ can be written in the form
		\[
		E^j X^k Y^l.
		\]
	\end{itemize}
Furthermore
	\be
	\lang E\rang\cap \lang X,\s Y \rang=\{\mathbbm{1}_3\},
	\ee
thus
	\be
	C(n,a,b)=\lang X,\s Y\rang\rtimes \lang E\rang.
	\ee
From theorem~\ref{AbelianSU3subgroups} we find 
	\be
	\lang X,\s Y\rang\cong\zed_m\times\zed_p,
	\ee
where
	\be\label{mdefinition}
	m=\max_{A\in\lang X,\s Y\rang}\mathrm{ord}(A)=\mathrm{ord}(X)=\mathrm{ord(Y)}=\mathrm{lcm}(\mathrm{ord}(\eta^a),\s \mathrm{ord}(\eta^b)).
	\ee
$\mathrm{lcm}(r,s)$ denotes the lowest common multiple of $r,s\in\mathbbm{N}$. Defining
	\be\label{pdefinition2}
	p:=\min\s\{k\in\{1,...,m\}\vert\s Y^k\in\lang X\rang\},
	\ee
we find
	\be
	\lang X,\s Y\rang=\{X^i Y^j\vert\enspace i=0,...,m-1;\enspace j=0,...,p-1\}\Rightarrow \lang X,\s Y\rang\cong \zed_m\times\zed_p. 
	\ee
For the sake of completeness we also want to find one possible choice of generators of $\zed_m$ and $\zed_p$. Applying the same argumentation as above on the definition
	\[
	q:=\min\s\{k\in\{1,...,m\}\vert\s X^k\in\lang Y\rang\},
	\]
we find $\lang X,\s Y\rang\cong\zed_m\times\zed_q\Rightarrow q=p$, and therefore
	\be\label{intersectionXY}
	\lang X\rang\cap\lang Y\rang=\lang X^p\rang=\lang Y^p\rang.
	\ee
Since $\mathrm{ord}(X^p)=\frac{m}{p}$ this leads to
		\be\label{tdefinition}
		\exists\s t\in \left\{1,...,\frac{m}{p}-1\right\}:\enspace X^{pt}=Y^p\Rightarrow (YX^{-t})^p=\bone_3,
		\ee
		but
		\[
		(YX^{-t})^a\neq \bone_3 \mbox{ for } a<p,
		\]
		because otherwise $Y^a\in\lang X\rang$ for $a<p$, which would be a contradiction to the definition~(\ref{pdefinition2}) of $p$. Therefore
		\be
		\lang YX^{-t} \rang\cong\zed_p
		\ee
Noticing furthermore that
		\be
		\lang X\rang\cap \lang YX^{-t} \rang=\{\bone_3\}
		\ee
we finally find
		\be
		\lang X,\s Y\rang=\lang X\rang\times\lang YX^{-t}\rang\cong\zed_m\times\zed_p.
		\ee
Now there are three cases:
	\begin{enumerate}
	\item $p=1$ $\Rightarrow Y\in \lang X\rang$ $\Rightarrow$ $\lang X,\s Y\rang=\lang X\rang$ $\Rightarrow$ $C(n,a,b)\cong \zed_m\rtimes\zed_3$.
	\item $p=m$ $\Rightarrow \lang X \rang\cap \lang Y \rang=\{\bone_3\}$\\
	      \phantom{$p=m$} $\Rightarrow$ $\lang X,\s Y\rang\cong \zed_m\times \zed_m$ $\Rightarrow$ $C(n,a,b)\cong (\zed_m\times\zed_m)\rtimes\zed_3\cong\Delta(3m^2)$.
	\item $p\in\{2,...,m-1\}$ $\Rightarrow C(n,a,b)\cong (\zed_m\times\zed_p)\rtimes\zed_3$.
	\end{enumerate}
$m$ is determined by equation~(\ref{mdefinition}) and from equations~(\ref{pdefinition2}) and~(\ref{tdefinition})  we can determine $p$ and $t$. One finds
	\be
	Y^p=X^{pt}\Rightarrow
	\begin{cases}
	p(b-at)\,\mathrm{mod}\, n=0,\enspace p\in\{1,...,m\},\mbox{ smallest possible}\\
	p(a+b(1+t))\,\mathrm{mod}\, n=0,\enspace t\in\{1,...,\frac{m}{p}-1\}
	\end{cases} 
	\ee
Let us summarize our results on the structure of the groups of type (C):
	\begin{itemize}
		\item $C(n,a,b)\cong (\zed_m\times\zed_p)\rtimes\zed_3,\mbox{ where}$
		\item $m=\mathrm{lcm}(\mathrm{ord}(\eta^a),\s \mathrm{ord}(\eta^b))$, and
		\item $\begin{cases}
			p(b-at)\,\mathrm{mod}\s n=0,\enspace p\in\{1,...,m\},\mbox{ smallest possible},\\
			p(a+b(1+t))\,\mathrm{mod}\s n=0,\enspace t\in\{1,...,\frac{m}{p}-1\}.
			\end{cases}$
		\item In terms of generators: $C(n,a,b)=(\lang X\rang\times\lang YX^{-t}\rang)\rtimes\lang E\rang$.
		\item $\mathrm{ord}(C(n,a,b))=3mp$.
		\item $p=1\Rightarrow C(n,a,b)\cong\zed_m\rtimes\zed_3$ ( $\rightarrow T_m$ for appropriate $m$ (see~\cite{BLW1,BLW2,Fairbairn2})). 
		\item $p=m\Rightarrow C(n,a,b)\cong\Delta(3m^3)\cong (\zed_m\times\zed_m)\rtimes\zed_3$.
	\end{itemize}
Note that (C)-groups can be direct products with $\zed_3$, i.e. the case
	\[
	C(n,a,b)\cong (\zed_{3x}\times\zed_y)\rtimes\zed_3 \cong ((\zed_x\times \zed_y)\rtimes\zed_3)\times\zed_3
	\]
is possible for some choices of $n,a,b$. Examples for this case are the groups
	\be
	C(6,1,1)\cong(\zed_6\times\zed_2)\rtimes\zed_3\cong((\zed_2\times\zed_2)\rtimes\zed_3)\times\zed_3\cong A_4\times\zed_3
	\ee
and the group $T_{21}$, which is described in~\cite{Fairbairn2}:
	\be
	T_{21}\cong C(21,1,4)\cong \zed_{21}\rtimes\zed_3\cong (\zed_7\rtimes\zed_3)\times\zed_3.
	\ee
As the last part of our investigation of the group series (C) we want to give an example for a (C)-group which neither belongs to the series $\Delta(3n^2)$, nor to the groups of type $T_n$. We already encountered an example in the group $C(6,1,1)$, but we also want to give an example for a ``new'' SU(3)-subgroup which is not just a direct product with an already well-known group. In~\cite{U3-512} all groups of order smaller 512 which possess a faithful three-dimensional irreducible representation (and are not isomorphic to a direct product with a cyclic group) have been listed. Among these groups $C(9,1,1)$ appears as the smallest (C)-group which is not classified as $T_n$ or $\Delta(3n^2)$. Using the tools we have developed in this section we immediately find
	\begin{itemize}
	\item $n=9$, $a=b=1$.
	\item $m=\mathrm{lcm}(\mathrm{ord}(\eta^a),\s\mathrm{ord}(\eta^b))=\mathrm{lcm}(9,9)=9$.
	\item The equations for $p$ and $t$ read
		 $\begin{cases}
			p(1-t)\,\mathrm{mod}\, 9=0,\enspace p\in\{1,...,9\},\mbox{ smallest possible},\\
			p(2+t)\,\mathrm{mod}\, 9=0,\enspace t\in\{1,...,\frac{9}{p}-1\}.
			\end{cases}$
		with the solution $p=3$, $t=1$.
	\item Thus
		\be
		C(9,1,1)\cong (\zed_9\times\zed_3)\rtimes\zed_3,
		\ee
		which coincides with the structure description given in~\cite{parattu-wingerter}, where the group is named by its SmallGroup number\footnote{See~\cite{U3-512,parattu-wingerter} for an explanation of the SmallGroup number (also called GAP ID in~\cite{parattu-wingerter}).} \textlbrackdbl 81,9\textrbrackdbl.
	\end{itemize}	

\section{On the structure of the SU(3)-subgroups of type (D)}
According to~\cite{miller} the groups of type (D) are generated by the matrices 
	\be
	E=\bmat
	0 & 1 & 0\\
	0 & 0 & 1\\
	1 & 0 & 0\\
	\emat,\quad
	F=F(n,a,b):=\bmat
	\eta^a & 0 & 0\\
	0 & \eta^b & 0\\
	0 & 0 & \eta^{-a-b}		
	\emat,
	\ee
($\eta=\exp(2\pi i/n),\quad n\in \mathbbm{N}\backslash\{0\},\quad a,b\in \{0,...,n-1\}$) of (C) and an additional generator $\widetilde{G}$ of the form
	\be
	\widetilde{G}=\bmat
	x & 0 & 0\\
	0 & 0 & y\\
	0 & z & 0
	\emat.
	\ee
The conditions $\mathrm{det}\s \widetilde{G}=1$ and $\mathrm{ord}(\widetilde{G})<\infty$ lead to~\cite{DTPOL}
	\be
	\widetilde{G}=\widetilde{G}(d,r,s):=\bmat
	\delta^r & 0 & 0\\
	0 & 0 & \delta^s\\
	0 & -\delta^{-r-s} & 0	
	\emat
	\ee
with
	\be
	\delta=\exp(2\pi i/d),\quad d\in \mathbbm{N}\backslash\{0\},\quad r,s\in \{0,...,d-1\}.
	\ee
Thus we define
	\be
	D(n,a,b;d,r,s):=\lang E,\s F(n,a,b),\s \widetilde{G}(d,r,s)\rang.	
 	\ee
For a better understanding of the structure of (D) it is helpful to reformulate the generators of the group. We define
	\be
	A:=\widetilde{G}^2=\bmat
	\delta^{2r} & 0 & 0\\
	0 & -\delta^{-r} & 0\\
	0 & 0 & -\delta^{-r}
	\emat,\quad G':=E^2 \widetilde{G}^2 E\widetilde{G}=\bmat
	-1 & 0 & 0\\
	0 & 0 & \delta^{2r+s}\\
	0 & \delta^{-(2r+s)} & 0
	\emat,
	\ee
which leads to
	\be
	D(n,a,b;d,r,s)=\lang A,\s E,\s F,\s G'\rang.
	\ee
Our first important observation is that, as in the case of (C), the subgroup $\mathcal{A}$ of all diagonal matrices is a normal subgroup of (D). This lets us hope to find a semidirect product structure as in the case of (C).

The action of the (non-diagonal) generators of (D) on any diagonal matrix is given by
	\be\label{S3action}
	\begin{array}{l}
	G'^{-1}\s \mathrm{diag}(a,b,c)\s G'=\mathrm{diag}(a,c,b),\\
	E^{-1}\s\mathrm{diag}(a,b,c)\s E=\mathrm{diag}(c,a,b),\\
	(EG')^{-1}\s \mathrm{diag}(a,b,c)\s EG'=\mathrm{diag}(c,b,a),\\
	E^{-2}\s \mathrm{diag}(a,b,c)\s E^2=\mathrm{diag}(b,c,a),\\
	(E^2 G')^{-1}\s\mathrm{diag}(a,b,c)E^2 G'=\mathrm{diag}(b,a,c).
	\end{array}
	\ee
This describes an $S_3$-action, which is well known from $\Delta(6n^2)$~\cite{D6n^2}, so the structure of (D) will be, though not identical in general, very similar to the structure
	\be
	\Delta(6n^2)\cong (\zed_n\times \zed_n)\rtimes S_3\cong ((\zed_n\times\zed_n)\rtimes\zed_3)\rtimes\zed_2
	\ee
of $\Delta(6n^2)$. Indeed
	\be
	\lang A,\s E,\s F,\s G'\rang\cong (\mathcal{A}\rtimes\lang E\rang)\rtimes \lang G'\rang\cong(\mathcal{A}\rtimes\zed_3)\rtimes\zed_2.
	\ee
In~\cite{ReviewGrimusLudl} it was shown that in fact
	\be
	(\mathcal{A}\rtimes\lang E\rang)\rtimes \lang G'\rang\cong \mathcal{A}\rtimes S_3.
	\ee
This can be illustrated by means of a similarity transformation~\cite{ReviewGrimusLudl}. We define
	\be
	T:=\mathrm{diag}(-\delta^{-2r-s},\,-\delta^{2r+s},\,1)
	\ee
and find
	\be
	\begin{split}
	& T^{-1}G'T=\bmat
	-1 & 0 & 0\\
	0 & 0 & -1\\
	0 & -1 & 0
	\emat=:G,\\
	& T^{-1}ET=
	\underbrace{\bmat
	\delta^{4r+2s} & 0 & 0\\
	0 & -\delta^{-2r-s} & 0\\
	0 & 0 & -\delta^{-2r-s}
	\emat}_{(EG'E)^2=:B\in\mathcal{A}}\bmat
	0 & 1 & 0\\
	0 & 0 & 1\\
	1 & 0 & 0
	\emat=BE,\\
	& T^{-1}\mathcal{A}T=\mathcal{A}.
	\end{split}
	\ee
Thus
	\be
	\lang A,\, E,\, F,\, G'\rang\cong\lang A,\, B,\, E,\, F,\, G\rang\cong\mathcal{A}\rtimes\lang E,\,G\rang,
	\ee
and due to $\lang E,\,G\rang\cong S_3$
	\be\label{Dsemidirectproduct}
	\lang A,\, B,\, E,\, F,\, G\rang\cong\mathcal{A}\rtimes S_3
	\ee
with $\lang A,\,B,\,F\rang\subset\mathcal{A}$.
For the explicit construction of $\mathcal{A}$ we refer the reader to appendix~\ref{mathcalA}.
Theorem~\ref{AbelianSU3subgroups} and equation~(\ref{Dsemidirectproduct}) lead to
	\be
	D(n,a,b;d,r,s)\cong\mathcal{A}\rtimes\lang E,\,G\rang\cong\mathcal{A}\rtimes S_3\cong(\zed_p\times\zed_q)\rtimes S_3,
	\ee
where $p$ and $q$ are functions of $n,a,b,d,r,s$.

While every group of type (C) can be interpreted as an irreducible representation of a group of type $\Delta(3n^2)$ (see equation~(\ref{CDelta3n^2})), a similar statement does \textit{not} hold for (D) and $\Delta(6n^2)$. In the following we will show that there is at least one SU(3)-subgroup of type (D), which cannot be interpreted as an irreducible representation of some $\Delta(6n^2)$.
In~\cite{U3-512} the following (D)-group has been found:
	\[
	D(9,1,1;2,1,1)\cong\mbox{\textlbrackdbl}162,14\mbox{\textrbrackdbl},
	\]	
which is of order 162. Since $C(9,1,1)$ is invariant under the action of the $\zed_2$-generator $G$, we find
	\be\label{D911211structure}
	D(9,1,1;2,1,1)\cong \underbrace{C(9,1,1)}_{\mathcal{A}\rtimes\lang E\rang}\rtimes\lang G\rang\cong ((\zed_9\times\zed_3)\rtimes\zed_3)\rtimes\zed_2\cong(\zed_9\times\zed_3)\rtimes S_3.
	\ee
Equation~(\ref{D911211structure}) suggests that there might be an irreducible three-dimensional representation $\mathcal{D}$ of $\Delta(6n^2)\cong(\zed_n\times\zed_n)\rtimes S_3$ such that $\mathcal{D}(\Delta(6n^2))\cong D(9,1,1;2,1,1)$. However, this is not the case, which we will show in the following. 
\\
If we can show, that there is no three-dimensional irreducible representation $\mathcal{D}$ of $\Delta(6n^2)$ with $\mathrm{ord}(\mathcal{D}(\Delta(6n^2)))=162$, we have found an example for a (D)-group that cannot be interpreted as an irreducible representation of some $\Delta(6n^2)$. This is indeed possible:

\begin{prop}\label{D6nnproposition}
Let $\mathcal{D}$ be a three-dimensional irreducible representation of $\Delta(6n^2)$, then
	\be
	\exists\s m\in \{1,...,n\}: \enspace\mathcal{D}(\Delta(6n^2))\cong\Delta(6m^2).
	\ee
The proof of this proposition can be found in appendix~\ref{D6nnproof}. Since $162/6=27$ is not a square number, we have proven that $D(9,1,1;2,1,1)$ cannot be interpreted as an irreducible representation of some $\Delta(6n^2)$. 
\end{prop}

\section{Conclusions}
In this work we tried to shine some light onto the hitherto not very well known series (C) and (D) of finite subgroups of SU(3). We were able to show that
	\[
	C(n,a,b)\cong (\zed_m\times\zed_p)\rtimes\zed_3
	\]
and we gave a method for the determination of $m$ and $p$ from the parameters $n,a,b$. We could also give a simple example for a (C)-group which is neither of the form $\Delta(3n^2)$ nor of the form $T_n$, thus showing that (C) contains some hitherto unclassified subgroups of SU(3).
\\
For the SU(3)-subgroups of type (D) we could determine the structure
	\[
	D(n,a,b;d,r,s)\cong(\zed_p\times\zed_q)\rtimes S_{3},
	\]
where $p$ and $q$ are functions of $n,a,b,d,r,s$. Since every (D)-group is a subgroup of some $\Delta(6m^2)\cong(\zed_m\times\zed_m)\rtimes S_3$ it is tempting to assume that the (D)-groups can be interpreted as irreducible representations of $\Delta(6m^2)$. However, by giving a counterexample, we could show that this is not the case in general. 

We hope that the analysis given here can lead us a small step closer towards the goal of the classification of all finite subgroups of SU(3). Furthermore we hope that some of the ``new'' types of finite subgroups of SU(3) discovered in this work can be useful for application in particle physics.

\subsection*{Acknowledgment}
The author wants to thank Roman Zwicky and Thomas Fischbacher for fruitful discussions and the University of Southampton, where the idea for this paper was born, for its hospitality. The author also wishes to thank Walter Grimus and the referees for their important and helpful comments.

\begin{appendix}

\section{Proof of theorem~\ref{AbelianSU3subgroups}}\label{proofappendix}
\textbf{\ref{AbelianSU3subgroups} Theorem.} Every finite Abelian subgroup $\mathcal{G}$ of SU(3) is isomorphic to $\zed_m\times\zed_n$, where 
	\be
	m=\max_{a\in \mathcal{G}}\mathrm{ord}(a)
	\ee
and $n$ is a divisor of $m$.

\begin{proof}
Since $\mathcal{G}$ is an Abelian group of $3\times 3$-matrices, we can choose a basis in which all group elements are diagonal. Then, due to $\mathrm{det}(a)=1\enspace\forall a\in\mathcal{G}$, all elements of $\mathcal{G}$ are of the form
	\be
	\bmat
	\alpha & 0 & 0\\
	0 & \beta & 0\\
	0 & 0 & \alpha^{\ast}\beta^{\ast}
	\emat,\quad \alpha,\beta\in \mathrm{U(1)}.
	\ee
Let
	\be\label{mdef}
	m:=\max_{a\in \mathcal{G}}\mathrm{ord}(a).
	\ee
Then $a^m=\bone_3\enspace\forall a\in\mathcal{G}$, which we will prove by contradiction. Suppose $\exists\s a\in\mathcal{G}:\enspace a^{m}\neq\bone_3$. $\Rightarrow$ $\mathrm{ord}(a)$ does not divide $m$. Let
	\be
	g:=\mathrm{gcd}(\mathrm{ord}(a),m)<\mathrm{ord}(a)
	\ee
denote the greatest common divisor of $\mathrm{ord}(a)$ and $m$. Then the group $\lang a^g\rang$ is a nontrivial cyclic group.
Let now $b$ be an element of $\mathcal{G}$ of order $m$. Since $\mathrm{ord}(a^g)$ and $m$ have no common divisor larger than 1 we find: 
	\be
	\lang a^g\rang\cap\lang b\rang=\{\bone_3\}.
	\ee
	\[
	\Rightarrow\mathrm{ord}(\underbrace{a^g b}_{\in\mathcal{G}})=\mathrm{ord}(a^g)\s \mathrm{ord}(b)=\underbrace{\dfrac{\mathrm{ord}(a)}{\mathrm{gcd}(\mathrm{ord}(a),m)}}_{>1}\times m>m\Rightarrow \mbox{contradiction to~(\ref{mdef}).}
	\]
Defining
	\be
	\mu:=\exp(2\pi i/m),
	\ee
every element of $\mathcal{G}$ has the form
	\be
	\bmat
	\mu^{c} & 0 & 0\\
	0 & \mu^{d} & 0\\
	0 & 0 & \mu^{-c-d}
	\emat,\quad c,d\in\{0,...,m-1\}.
	\ee
Thus $\mathcal{G}$ is a subgroup of
	\[
	\left\langle\left\langle
	\bmat
	\mu & 0 & 0\\
	0 & 1 & 0\\
	0 & 0 & \mu^{\ast}
	\emat,\enspace
	\bmat
	1 & 0 & 0\\
	0 & \mu & 0\\
	0 & 0 & \mu^{\ast}
	\emat
	\right\rangle\right\rangle\cong\zed_m\times\zed_m.
	\]
Let
	\be
	m=p_1^{k_1}p_2^{k_2}\cdots p_j^{k_j}
	\ee
be the prime factorization of $m$ ($p_1,\dots, p_j$ are the distinct prime factors of $m$). Then every Abelian group
of order $m$ is the direct product of $j$ Abelian groups $A_i$ of order $p_i^{k_i}$ (see e.g. theorem~3.3.1 of~\cite{Hall}). Since
every subgroup of a cyclic group is cyclic (see e.g. theorem~3.1.1 of~\cite{Hall}), in the case of $\zed_m$ all $A_i$ are cyclic and we find
	\be\label{zedmfactorization}
	\zed_m\cong \zed_{p_1^{k_1}}\times\zed_{p_2^{k_2}}\times\cdots\times\zed_{p_j^{k_j}},
	\ee
and thus
	\be
	\zed_m\times\zed_m\cong(\zed_{p_1^{k_1}}\times\zed_{p_1^{k_1}})\times\cdots
\times(\zed_{p_j^{k_j}}\times\zed_{p_j^{k_j}}).
	\ee
Next we use the fact that if $p$ is prime, every subgroup of $\zed_{p^{e_1}}\times\cdots\times\zed_{p^{e_s}}$
is of the form $\zed_{p^{f_1}}\times\cdots\times\zed_{p^{f_s}}$ with $0\leq f_i\leq e_i$ (for a proof
of this statement see e.g. theorem~3.3.3 of~\cite{Hall}). Consequently, every subgroup $\mathcal{G}$ of $\zed_m\times\zed_m$
has the form
	\be\label{Gfactorization}
	\mathcal{G}\cong(\zed_{p_1^{r_1}}\times\zed_{p_1^{n_1}})\times\cdots
\times(\zed_{p_j^{r_j}}\times\zed_{p_j^{n_j}})\cong \zed_{r}\times\zed_{n},
	\ee
where $0\leq n_i\leq r_i\leq k_i$ and $r:=\prod_j p_j^{r_j}$,\,$n:=\prod_j p_j^{n_j}$.
Without loss of generality we have assumed $n_i\leq r_i$, which implies $1\leq n\leq r\leq m$.

$\zed_m$ is a subgroup of $\mathcal{G}$, because by definition~(\ref{mdef}) there exists at least
one element of order $m$ in $\mathcal{G}$. Therefore---from equation~(\ref{Gfactorization})---we
conclude $r=m$. By construction $n$ is a divisor of $r$, which completes the proof.
\end{proof}

\section{The group $\mathcal{A}$ of diagonal matrices in (D)}\label{mathcalA}
In this appendix we want to construct a generating set of the invariant subgroup $\mathcal{A}$ of all diagonal matrices in (D). The generators of $\mathcal{A}$ are
$A$, $B$ and $F$ as well as the actions of $E$ and $G$ on $A$, $B$ and $F$ (see equation~(\ref{S3action})).
	\be
	\begin{split}
	\mathcal{A}=
	\lang & A,\s B,\s F,\\
	& G^{-1}AG,\s E^{-1}AE,\s (EG)^{-1}AEG,\s E^{-2}AE^2,\s (E^2 G)^{-1}AE^2 G,\\
	& G^{-1}BG,\s E^{-1}BE,\s (EG)^{-1}BEG,\s E^{-2}BE^2,\s (E^2 G)^{-1}BE^2 G,\\
	& G^{-1}FG,\s E^{-1}FE,\s (EG)^{-1}FEG,\s E^{-2}FE^2,\s (E^2 G)^{-1}FE^2 G\rang.\\
	\end{split}
	\ee
For any diagonal phase matrix $D$ of determinant 1 (and thus for any element of $\mathcal{A}$) we have
	\be\label{redundantgenerators}
	\begin{split}
	& E^{-2}DE^2=D^{-1}(E^{-1}DE)^{-1}\quad\mbox{and}\\
	& (E^2G)^{-1}D(E^2G)=(G^{-1}DG)^{-1}[(EG)^{-1}D(EG)]^{-1},
	\end{split}
	\ee
therefore generators of the form~(\ref{redundantgenerators}) with $D=A,\,B,\,F$ are redundant. Using furthermore
	\be
	G^{-1}DG=D\quad\mbox{and}\quad(EG)^{-1}D(EG)=D^{-1}(E^{-1}DE)^{-1}
	\ee
for $D=A,\,B$, we end up with
	\be
	\begin{split}
	\mathcal{A}=
	\lang & A,\, E^{-1}AE,\\
	& B,\, E^{-1}BE,\\
	& F,\, E^{-1}FE,\, G^{-1}FG,\, (EG)^{-1}FEG\rang.\\
	\end{split}
	\ee

\section{Proof of proposition~\ref{D6nnproposition}}\label{D6nnproof}

\textbf{\ref{D6nnproposition} Proposition.}
Let $\mathcal{D}$ be a three-dimensional irreducible representation of $\Delta(6n^2)$, then
	\be
	\exists\s m\in \{1,...,n\}: \enspace\mathcal{D}(\Delta(6n^2))\cong\Delta(6m^2).
	\ee

\begin{proof}
Following~\cite{D6n^2} equation~(\ref{D6nnpresentation}) comprises a presentation of $\Delta(6n^2)$ in terms of four generators:
	\be\label{D6nnpresentation}
	\begin{array}{ll}
	P^3=Q^2=(PQ)^2=\bone &\quad S_3-\mbox{presentation}\\
	R^n=S^n=\bone,\enspace RS=SR &\quad \zed_n\times\zed_n-\mbox{presentation}\\
	PRP^{-1}=R^{-1}S^{-1},\enspace PSP^{-1}=R &\quad \mbox{action of } S_3\\
	QRQ^{-1}=S^{-1},\enspace QSQ^{-1}=R^{-1} & \quad \mbox{on } \zed_n\times\zed_n
	\end{array}	
	\ee
There are only two types of three-dimensional irreducible representations of $\Delta(6n^2)$, namely~\cite{D6n^2}
	\[
	\begin{split}
	& \textbf{\underline{3}}_{1(l)}: 
	P\mapsto
	\begin{pmatrix}
	0 & 1 & 0\\
	0 & 0 & 1\\
	1 & 0 & 0
	\end{pmatrix},\quad Q\mapsto	
	\begin{pmatrix}
	0 & 0 & 1\\
	0 & 1 & 0\\
	1 & 0 & 0
	\end{pmatrix},\quad R\mapsto 
	\begin{pmatrix}
	\eta^l & 0 & 0\\
	0 & \eta^{-l} & 0\\
	0 & 0 & 1
	\end{pmatrix},\quad S\mapsto
	\begin{pmatrix}
	1 & 0 & 0\\
	0 & \eta^{l} & 0\\
	0 & 0 & \eta^{-l}
	\end{pmatrix}\\
	& \textbf{\underline{3}}_{2(l)}: 
	P\mapsto
	\begin{pmatrix}
	0 & 1 & 0\\
	0 & 0 & 1\\
	1 & 0 & 0
	\end{pmatrix},\quad Q\mapsto	
	\begin{pmatrix}
	0 & 0 & -1\\
	0 & -1 & 0\\
	-1 & 0 & 0
	\end{pmatrix},\quad R\mapsto 
	\begin{pmatrix}
	\eta^l & 0 & 0\\
	0 & \eta^{-l} & 0\\
	0 & 0 & 1
	\end{pmatrix},\quad S\mapsto
	\begin{pmatrix}
	1 & 0 & 0\\
	0 & \eta^{l} & 0\\
	0 & 0 & \eta^{-l}
	\end{pmatrix}.
	\end{split}
	\]
$\eta:=\exp(2\pi i/n)$, $n\in\mathbbm{N}\backslash\{0,1\}$, $l\in\{1,...,n-1\}$.

The matrix groups defined by the irreducible representations given above fulfill the presentation (\ref{D6nnpresentation}) with $n$ replaced by
	\be
	m:=\mathrm{ord}(\eta^l),
	\ee
and thus
	\be
	\mathcal{D}(\Delta(6n^2))\cong\Delta(6m^2)
	\ee
for all three-dimensional irreducible representations of $\Delta(6n^2)$.
\end{proof}

\end{appendix}

\end{document}